\documentclass{an}
\usepackage{graphicx}
\usepackage{times}
\Volume{01}              
\Year{2002}              
\Month{10}               
\Pagespan{223}{227}      

\begin{document}

\def\cm{{\rm\thinspace cm}}
\def\erg{{\rm\thinspace erg}}
\def\s{{\rm\thinspace s}}

\def\ergpspcmsq{\hbox{$\erg\cm^{-2}\s^{-1}\,$}}
\def\ergps{\hbox{$\erg\s^{-1}\,$}}
\def\pcmsq{\hbox{$\cm^{-2}\,$}}

\lhead[\thepage]{C.S. Crawford, P. Gandhi \& A.C. Fabian : Optical and
near-infrared observations of hard serendipitous {\sl Chandra} sources}
\rhead[Astron. Nachr./AN~{\bf XXX} (2003) X]{\thepage}
\headnote{Astron. Nachr./AN {\bf 32X} (200X) X, XXX--XXX}

\title{Optical and near-infrared observations of hard serendipitous
{\sl Chandra} sources}

\author{C.S. Crawford, P. Gandhi \and  A.C. Fabian}
\institute{
Institute of Astronomy, Madingley Road, Cambridge CB3 0HA, UK}

\date{Received {\it date will be inserted by the editor}; 
accepted {\it date will be inserted by the editor}} 

\abstract{
We have been carrying out a successful observational programme
targeted at finding the highly obscured quasars that are thought to be
the main contributors to the hard X-ray background. Out of 56 sources
so far studied with optical and near-infrared imaging and
spectroscopy, we have found three definite and a further twelve
possible Type II quasars. Few sources show significant line emission,
suggesting that the line photons are depleted by the large columns of
obscuring matter. The redshift distribution of our sources shows a
distinct peak at $z\sim1$. The broad-band colours and magnitudes
of the optical/near-infrared counterparts indicate that the light in
these bands is dominated by the continuum of a massive bright galaxy. 
}

\correspondence{csc@ast.cam.ac.uk}

\maketitle

\section{Introduction}

Standard models synthesize the hard (2-10keV) X-ray background
(hereafter HXRB) as the integrated emission from a population of
obscured active galactic nuclei (AGN); the hard spectral slope is
produced by the presence of large amounts of intrinsic obscuration
that preferentially erode the soft photons (Setti \& Woltjer 1989;
Comastri et al 1995; Wilman \& Fabian 1999). The models
suggest that the sources with the most accretion activity are hidden
as highly obscured AGN at a range of both absorbing columns and
redshift.Thus a large population of powerful active objects at high
redshifts is expected to be \lq missing' from conventional
(optically-based) surveys, and this regime may be where most of the
growth of massive black holes occurs. 

We report here on our work in progress aiming to find such absorbed
powerful sources in sufficient numbers to determine the importance of
their contribution to the hard X-ray background.

\section{The Sample}

In contrast to deep narrow surveys (eg Brandt et al 2001; Hasinger et
al 2001; Giacconi et al 2002), we have undertaken a wide-area search,
targeted towards finding \lq type II' highly obscured quasars (ie
objects with intrinsic L$_{\rm X}>2\times10^{44}$\ergps). We have
selected X-ray sources from those found serendipitously in the field
of our own and archival {\sl Chandra} ACIS-S observations, most of
which are of clusters of galaxies (out to $z\sim0.5$). Whilst this
means that the central area of the most sensitive ACIS chip is taken
up by the soft thermal emission from the intracluster medium, the
cores of clusters are also popular targets for optical imaging, so
deep multi-wavelength images are already available from telescope
archives. In addition, gravitational lensing by the foreground cluster
can boost distant sources into observability (eg in A2390, Cowie et al
2001; Crawford et al 2002). [Contamination by sources within the
cluster itself has not proven to be a problem, as we have only found
two sources to have a redshift consistent with that of the cluster.]

We have compiled a sample so far of 341 sources with more than 10 net
counts taken from 11 fields.  131 of these sources are hard, with a
significant number of their counts emerging above 2~keV. At the
typical 10-20~ksec exposure times of our {\sl Chandra} observations
this means we are detecting sources with fluxes of $10^{-13}$ to
$10^{-15}$\ergpspcmsq, straddling the \lq knee' in the hard source
counts log$N$-log$S$ distribution (Rosati et al 2002). Thus in this regime
we are selecting the sources which are the dominant flux contributors
to the HXRB. Deeper surveys do {\sl not} find obscured quasars in
great numbers (only two have been reported: Norman et al 2002; Stern
et al 2002), though they do find many Seyfert IIs (with intrinsic
L$_{\rm X}<2\times10^{44}$\ergps). Due to the flattening in source
counts, we should be able to detect more of the powerful sources by
studying a larger number of less deep fields than would be found in a
single deep field.

\section{Observations} 

A preliminary investigation of 31 serendipitous sources in the A~2390
cluster (Crawford et al 2002) showed that the optical spectra of the
X-ray hard sources fell into two broad classes: a bright population at
low redshift ($z<0.3$) with strong but narrow emission lines,
consistent with Compton-thick Seyfert II galaxies; and a distant
($z>1$) faint population of hard sources with very weak, or no
emission lines. Two of the sources were found to be Type II quasars
with obscuring columns in excess of N$_{\rm H}>10^{23}$\pcmsq, and an
intrinsic bolometric luminosity $>10^{45}$\ergps; in both cases strong
gravitational lensing by the foreground cluster enhanced their
detection.

As we expect our target sources to be highly redshifted ($z>1$) and/or
absorbed, we moved to the near-infrared to seek counterparts for the
hard X-ray sources from our full sample that are very faint (or
absent) in the optical band. We imaged the fields of the sources in
the $J$, $H$ and $K$ bands, using both UKIRT and VLT. We have
near-infrared detections of 56 sources to $K<$20, 37 of which are hard
(the soft sources were included as they happened to be in the field of
view of some of the hard sources). Nearly all of the optically-faint
sources are relatively bright and readily detected in the
near-infrared with a median $K$ magnitude of $\sim$18 (Crawford et al
2001; Gandhi et al 2002a). Most are clearly resolved, and a few have a
double (possibly interacting) morphology (eg Fig~1).

\begin{figure}
{\includegraphics[angle=0,width=5cm] {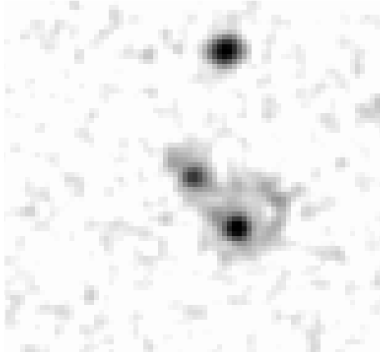}}
\caption{
10$''\times10''$ $K$-band image of the $K=19$ source MS2137\_4 taken
with the VLT, showing its curious double morphology with diffuse
(possibly tidal) structure. The hard X-ray source is associated with
the NE component of the pair, but both have a very red $B-K$ colour.
The estimated redshift is $z_{phot}\sim3.3$. }
\end{figure}

We also obtained broad-band near-infrared spectra of 20 of these hard
sources, again using both UKIRT and VLT (Crawford et al 2001; Gandhi
et al 2002a). Only 3 sources had strong detectable line emission, and
only one of these had an unambiguous redshift identification, from a
clear detection of H$\alpha$, [NII] and [SII] at a redshift of 2.176
(Fig~2). Most of the sources had a flat continuum spectrum, showing no
significant emission lines to a best limiting equivalent width of
20\AA.

\begin{figure}
\includegraphics[angle=270,width=8.5cm]{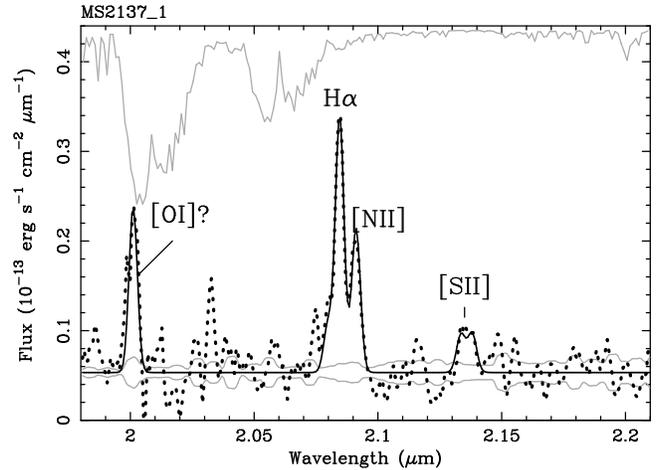}
\caption{ ISAAC $K$-band spectrum of the source MS2137\_1 at a
redshift of $z=2.176$. The data are shown as bold dots, the model
fitted to the emission lines as a solid line and the 1$\sigma$ errors
due to the sky above and below a constant fitted continuum as the
faint grey lines. Sky absorption has been rescaled to be plotted as
the faint grey line at the top of the figure. The spectrum has been
smoothed over 5 adjacent pixels.}
\end{figure}

The lack of identifiable emission line or absorption features in our
near-infrared spectra made it expedient to use photometric redshift
estimation to characterize most of our source population. We used the
publicly-available code HYPERZ (Bolzonella et al 2000) which is based
on fitting template continua to broad-band photometric fluxes in as
many bands as possible. A check of the derived $z_{phot}$ against a
measured $z_{spec}$ (where available) in our sample found that they
agree well for the hard sources; but less well for the soft sources,
where there may be a larger contribution to the broad band fluxes from
the (variable) quasar continuum and emission lines.

\section{Results and Discussion} 

To investigate the surprising lack of significant spectral features in
our near-infrared spectra of hard sources, we predicted the expected
broad line strengths based on the observed correlation between the
2-10~keV X-ray luminosity and the luminosity in the broad H$\alpha$
emission line in emission-line AGN (Ward et al 1988). In all seven
targets where we have a redshift identification (or a good photometric
redshift estimate), the predicted line emission is sufficiently strong
that we expect to have observed broad lines above a 2$\sigma$
threshold, and in at least four cases, above a 3$\sigma$ threshold.
One source (MS2137\_1; Fig~2) that is predicted to have a strong broad
H$\alpha$ line does not show this component, although the narrow
component is clearly visible. Alternatively, if the estimated redshift
is not known (or very wrong), it is still very unlikely that our
bandpasses systematically missed {\sl any} (either broad or narrow)
emission lines in so many sources; there are few regions in redshift
space where no redshifted lines are observed in either (or both) of
the observed bandpasses (we expect Hydrogen or MgII lines to be
visible for most of $0<z<5$). We thus infer that our sources are
dominated by the continuum light from the host galaxy in the
near-infrared, with some mechanism inhibiting or obscuring the line
emission. The lines are most likely depleted due to the large columns
of obscuring gas and associated dust intrinsic to the sources.

The redshift distribution of the sources studied in detail so far
(Fig~3) shows a distinct peak at $z\sim1$, with a small tail of
sources extending to higher redshift, in broad agreement with findings
from follow-up on the deep fields (Rosati et al 2002; Hasinger 2002;
Brandt et al 2002). Both the hard and soft sources are similarly
spread in redshift. Additional evidence that the optical and
near-infrared light from the counterparts of hard sources is dominated
by the host galaxy comes from their broad-band colours, which are redder
than those of an unobscured QSO. Some are extremely red objects, with
$R-K>5$. The hard sources in our sample cover the brighter region of
$K-z$ space, most having luminosities brighter than $L^*$ at all
redshifts. This suggests that they are located in very massive and
bright galaxies, exactly those assumed to host some of the most
massive black holes (Magorrian et al 1998; Merrit \& Ferrarese 2001).

\begin{figure}
\includegraphics[angle=0,width=8.5cm]{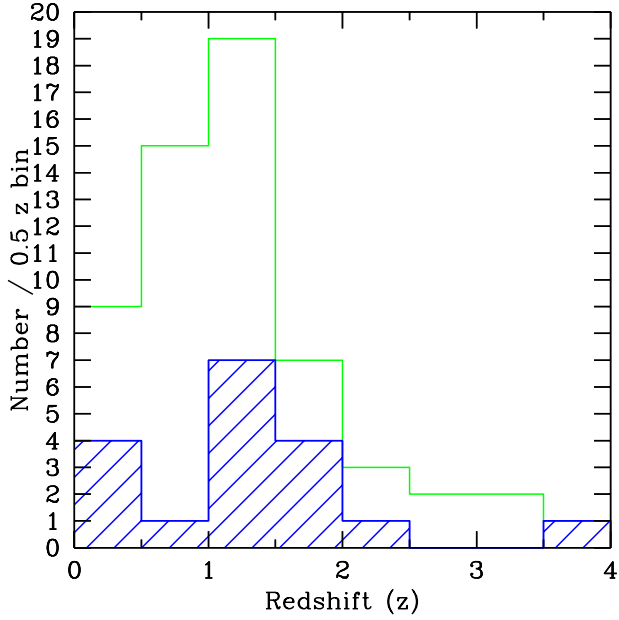}
\caption{
Redshift histogram for 18 sources with spectroscopic measurements
(hatched regions) and 40 additional sources with photometric redshifts
only (clear outlined region).}
\end{figure}

The bulk of the hard objects in our sample have X-ray column densities
which classify them as type II AGN. We have been successful in
identifying three {\sl bona fide} Type II QSO (two in the field of
A~2390 Crawford et al 2002; the other is shown in Fig~4 and in Gandhi
et al 2002b), with high intrinsic luminosities (L$_{\rm
X}>2\times10^{44}$\ergps) and obscuring column densities (N$_{\rm
H}>10^{23}$\pcmsq).

\begin{figure}
{\includegraphics[angle=0,width=5cm] {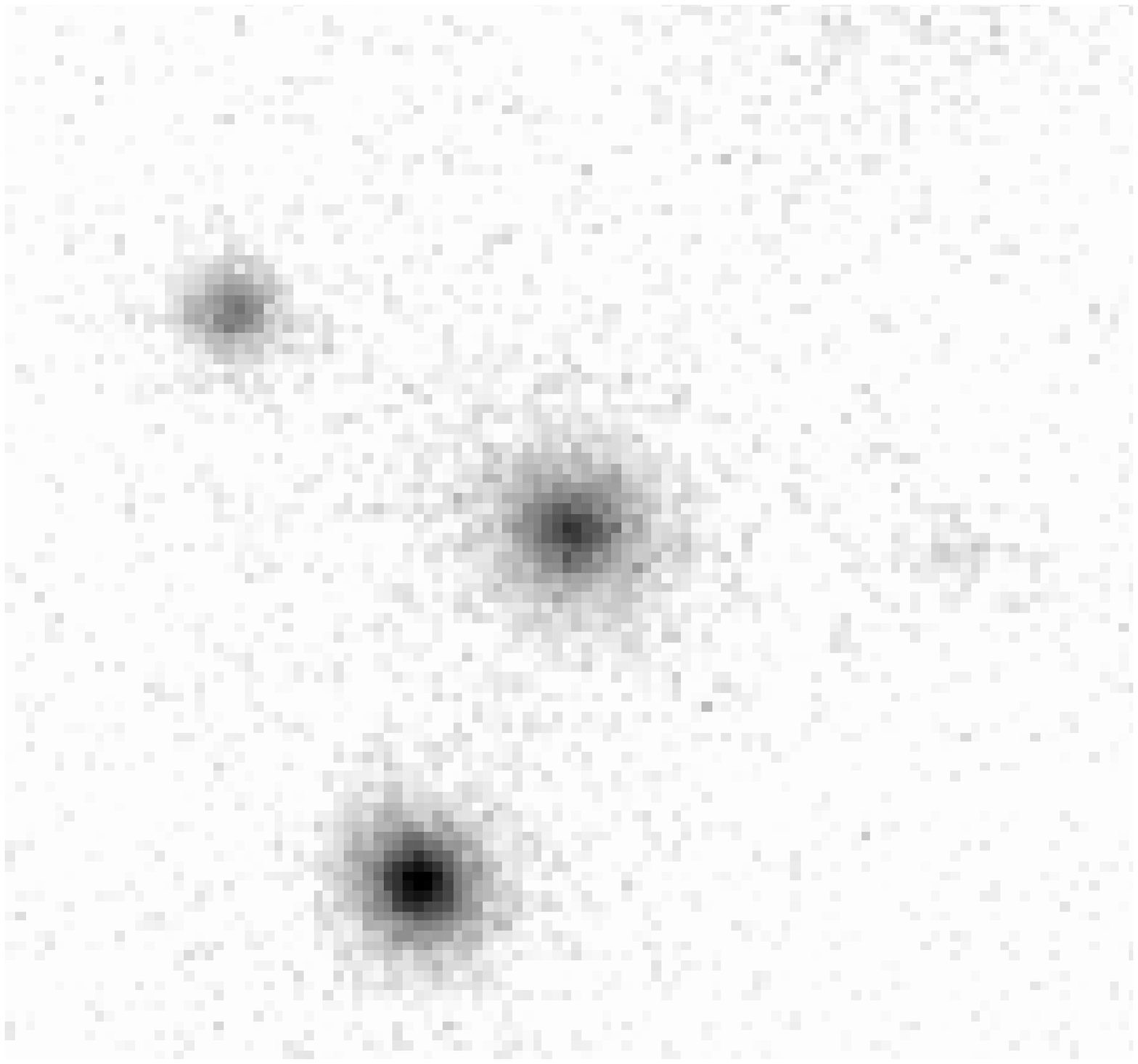}} 
{\includegraphics[angle=270,width=8.5cm] {a963_15_x1.ps}} 
{\includegraphics[angle=270,width=8.5cm] {a963_15_x2.ps}} 
\caption{
(Top) UKIRT UFTI $K$-band image of the $K=16.5$ host galaxy of A963\_15, which has a
photometric redshift of $z\sim0.56$. The image is 10 arcsec on a side,
and the target is the central object.\newline
(Middle) {\sl Chandra} spectrum showing A963\_15 to be a very hard
source, with a total flux of 1.8$\times10^{-14}$\ergpspcmsq. The data
(solid circles in light grey) have been fit to a power-law model (solid line) with a
$\Gamma=2$ and a 6.4~keV Fe K$\alpha$ line at $z=0.54$. The fitted
intrinsic absorption is N$_{\rm H}=1.1\times10^{24}$\pcmsq. The dotted
line shows the power-law model if it were affected only by Galactic
line-of-sight obscuration. \newline
(Bottom) 68, 95 and 99 per cent confidence intervals for the power-law fit, showing that
the evidence for a high absorbing column is robust for all reasonable values of $\Gamma$. }
\end{figure}

We have a further six sources with hard X-ray colours and an X-ray
luminosity of L$_{\rm X}>10^{45}$\ergps -- assuming they lie at a
redshift equal to or greater than their $z_{phot}$ or $z_{spec}$, and
have a power-law spectrum with $\Gamma=1.4$ -- or a total of twelve
such potential Type II quasars, with L$_{\rm X}>3\times10^{44}$\ergps
(eg in Fig~5). These luminosities are derived assuming only Galactic
absorption; obviously the hard X-ray colours suggest the further
presence of large amounts of intrinsic absorption that will increase
the inferred luminosity. Assuming a steeper slope $\Gamma$ will only
act to increase the intrinsic column density required. Aside from
A963\_15 (Fig~4), the photometric redshifts of these putative Type II
quasars range over $1.3<z<3.4$, and their optical magnitudes range
from $I\sim19.5$ down to $I>23.6$. Detailed follow-up observations of
such sources, especially given the lack of emission line features in
their spectra, will require large (8m) telescopes.

\begin{figure}
{\includegraphics[angle=0,width=5cm] {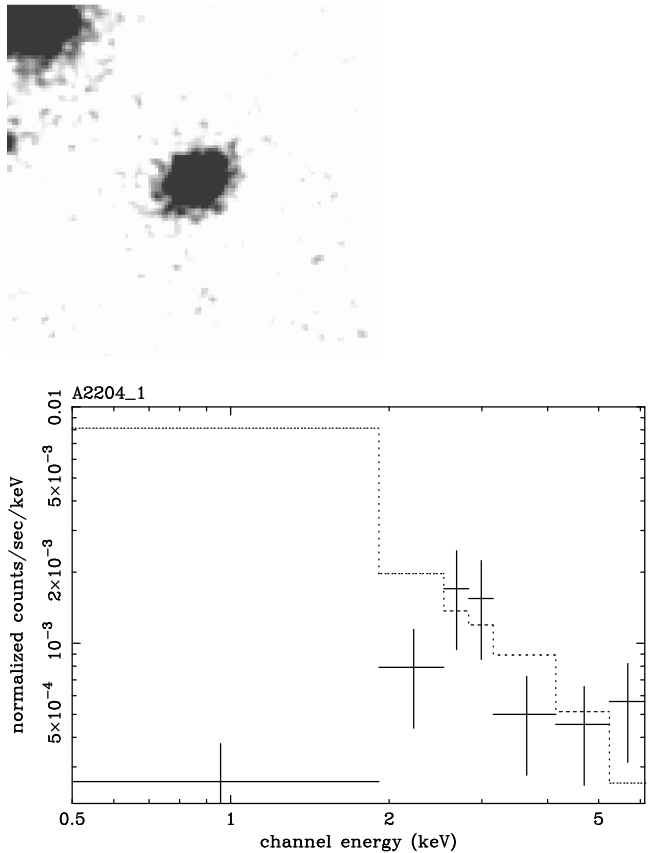}}
{\includegraphics[angle=270,width=8.5cm] {a2204611_grp5_gal.ps}}
\caption{
A $K$-band image (10 arcsec on a side) of the $K=16.4$ host galaxy
associated with A2204\_1 (top) which has  $z_{phot}\sim0.5$.
The {\sl Chandra} X-ray spectrum (bottom) clearly
demonstrates the hard nature of the source, with the least counts being
detected in the soft band, where a $\Gamma=2$ power law model with
only Galactic absorption would predict the most flux (dotted line).
}
\end{figure}

\section {Conclusions}

{\sl Chandra} has been successful in detecting highly obscured AGN ,
even in relatively short exposures. We have a well-selected sample of
three definite, and a further twelve possible, type II quasars. This
number should increase as we observe further optically-faint,
serendipitous hard sources from our sample.

\end{document}